\documentclass{ws-procs9x6}

\usepackage{graphicx,color}
\usepackage{epic,eepic}
\usepackage{amsmath,amssymb}
\usepackage{booktabs,colortbl}
\usepackage{tabularx}

\def\bra{\langle}
\def\ket{\rangle}

\def\qbar{\overline{{\rm q}}}

\def\sbar{\overline{{\rm s}}}

\def\threebar{\overline{{\mathbf{3}}}}

\def\calO{{\mathcal{O}}}
\def\OFS{\bra\calO_{FS}\ket}

\def\Voge{{V_{\rm OGE}}}
\def\Vobe{{V_{\rm PS}}}
\def\Vconf{{V_{\rm conf}}}
\def\Vconflam{{V_{\rm conf}^{(\lambda)}}}
\def\Vconfone{{V_{\rm conf}^{(1)}}}
\def\aconf{{a_{\rm conf}}}

\def\lamilamj{(\lambda_i\cdot\lambda_j)}
\def\sigisigj{(\sigma_i\cdot\sigma_j)}

\def\fifj{(f_i\cdot f_j)}

\def \mbf(#1){\mbox{\boldmath{$#1$}}}

\def\vecxi{\mbf(\xi)}
\def \veceta{\mbf(\eta)}
\def \vecr{\mbf(r)}
\def \vecR{\mbf(R)}
\def\talpha{{\tilde \alpha}}
\def\tbeta{{\tilde \beta}}
\def\tgamma{{\tilde \gamma}}
\def\tdelta{{\tilde \delta}}

\def\MeV{{\footnotesize{[MeV]}}}
\def\OGE{{\footnotesize{OGE}}}

\def\ohalfp{\smash{${1\over 2}^+$}}
\def\ohalfm{\smash{${1\over 2}^-$}}
\def\thalfm{\smash{${3\over 2}^-$}}

\newcommand{\FRAC}[2]{\leavevmode\kern.1em
  \raise.5ex\hbox{\the\scriptfont0 #1}\kern-.1em
  /\kern-.15em\lower.25ex\hbox{\the\scriptfont0 #2}}

\definecolor{Mygray}{gray}{0.5}
\definecolor{Myblue}{rgb}{0,0.3,1}
\definecolor{Myred}{rgb}{1,0,0}
\definecolor{Mygreen}{rgb}{0,1,1}

\def\Mp#1#2#3#4#5#6#7{%
\ifnum#1=0%
  \ifcase#3 \def\Tpt{1500}
  \or\def\Tpt{1500}
  \or\def\Tpt{1500}
  \or\def\Tpt{1650}
  \or\def\Tpt{1650}
  \or\def\Tpt{1800}
  \else
  \def\Tpt{1500}
  \fi
\fi%
\ifnum#1=1%
  \ifcase#3 \def\Tpt{2000}
  \or\def\Tpt{2000}
  \or\def\Tpt{2000}
  \or\def\Tpt{2150}
  \or\def\Tpt{2150}
  \or\def\Tpt{2300}
  \else
  \def\Tpt{2000}
  \fi
\fi%
\ifnum#1=2%
  \ifcase#3 \def\Tpt{2500}
  \or\def\Tpt{2500}
  \or\def\Tpt{2500}
  \or\def\Tpt{2650}
  \or\def\Tpt{2650}
  \or\def\Tpt{2800}
  \else
  \def\Tpt{2500}
  \fi
\fi%
\put(\Tpt,#4){{\makebox(0,0){\rule{6.8pt}{0.5pt}\rule{6.2pt}{0pt}}}}%
\put(\Tpt,#5){{\makebox(0,0){\rule{6.3pt}{0pt}\rule{6.7pt}{1.5pt}}}}%
\put(\Tpt,#4){{\thinlines\line(0,#6){#7}}}%
}

\def\Mpx#1#2#3#4#5#6#7{%
\ifnum#1=0%
  \ifcase#3 \def\Tpt{1500}
  \or\def\Tpt{1500}
  \or\def\Tpt{1500}
  \or\def\Tpt{1650}
  \or\def\Tpt{1650}
  \or\def\Tpt{1800}
  \else
  \def\Tpt{1500}
  \fi
\fi%
\ifnum#1=1%
  \ifcase#3 \def\Tpt{2000}
  \or\def\Tpt{2000}
  \or\def\Tpt{2000}
  \or\def\Tpt{2150}
  \or\def\Tpt{2150}
  \or\def\Tpt{2300}
  \else
  \def\Tpt{2000}
  \fi
\fi%
\ifnum#1=2%
  \ifcase#3 \def\Tpt{2480}
  \or\def\Tpt{2480}
  \or\def\Tpt{2480}
  \or\def\Tpt{2630}
  \or\def\Tpt{2630}
  \or\def\Tpt{2780}
  \else
  \def\Tpt{2500}
  \fi
\fi%
\put(\Tpt,#4){{\makebox(0,0){\rule{6.8pt}{0.5pt}\rule{6.2pt}{0pt}}}}%
\put(\Tpt,#5){{\makebox(0,0){\rule{6.3pt}{0pt}\rule{6.7pt}{1.5pt}}}}%
\put(\Tpt,#4){{\thinlines\line(0,#6){#7}}}%
}

\def\Ms#1#2#3#4#5#6#7{%
\ifnum#1=0%
  \ifcase#3 \def\Tpt{0}
  \or\def\Tpt{0}
  \or\def\Tpt{0}
  \or\def\Tpt{150}
  \or\def\Tpt{150}
  \or\def\Tpt{300}
  \else
  \def\Tpt{0}
  \fi
\fi%
\ifnum#1=1%
  \ifcase#3 \def\Tpt{500}
  \or\def\Tpt{500}
  \or\def\Tpt{500}
  \or\def\Tpt{650}
  \or\def\Tpt{650}
  \or\def\Tpt{800}
  \else
  \def\Tpt{500}
  \fi
\fi%
\ifnum#1=2%
  \ifcase#3 \def\Tpt{1000}
  \or\def\Tpt{1000}
  \or\def\Tpt{1000}
  \or\def\Tpt{1150}
  \or\def\Tpt{1150}
  \or\def\Tpt{1300}
  \else
  \def\Tpt{1000}
  \fi
\fi%
\put(\Tpt,#4){{\makebox(0,0){\rule{6.8pt}{0.5pt}\rule{6.2pt}{0pt}}}}%
\put(\Tpt,#5){{\makebox(0,0){\rule{6.3pt}{0pt}\rule{6.7pt}{1.5pt}}}}%
\put(\Tpt,#4){{\thinlines\line(0,#6){#7}}}%
}

\def\Msx#1#2#3#4#5#6#7{%
\ifnum#1=0%
  \ifcase#3 \def\Tpt{-20}
  \or\def\Tpt{-20}
  \or\def\Tpt{-20}
  \or\def\Tpt{130}
  \or\def\Tpt{130}
  \or\def\Tpt{280}
  \else
  \def\Tpt{0}
  \fi
\fi%
\ifnum#1=1%
  \ifcase#3 \def\Tpt{480}
  \or\def\Tpt{480}
  \or\def\Tpt{480}
  \or\def\Tpt{630}
  \or\def\Tpt{630}
  \or\def\Tpt{780}
  \else
  \def\Tpt{500}
  \fi
\fi%
\ifnum#1=2%
  \ifcase#3 \def\Tpt{980}
  \or\def\Tpt{980}
  \or\def\Tpt{980}
  \or\def\Tpt{1130}
  \or\def\Tpt{1130}
  \or\def\Tpt{1280}
  \else
  \def\Tpt{1000}
  \fi
\fi%
\put(\Tpt,#4){{\makebox(0,0){\rule{6.8pt}{0.5pt}\rule{6.2pt}{0pt}}}}%
\put(\Tpt,#5){{\makebox(0,0){\rule{6.3pt}{0pt}\rule{6.7pt}{1.5pt}}}}%
\put(\Tpt,#4){{\thinlines\line(0,#6){#7}}}%
}

\def\Waku#1#2#3{%
\put(-200,#2){\framebox(3200,#3)[t]{%
\hfill \rule{0pt}{22pt}\framebox{\sf #1}~~}}%
\put(-100,#2){\makebox(100,-60)[t]{\small $(2S)^P$=$1^-\,3^-\,5^-$}}%
\put(650,#2){\makebox(0,-60)[t]{\small $1^-\,3^-\,5^-$}}%
\put(1150,#2){\makebox(0,-60)[t]{\small $1^-\,3^-\,5^-$}}%
\put(1650,#2){\makebox(0,-60)[t]{\small $1^+\,3^+\,5^+$}}%
\put(2150,#2){\makebox(0,-60)[t]{\small $1^+\,3^+\,5^+$}}%
\put(2650,#2){\makebox(0,-60)[t]{\small $1^+\,3^+\,5^+$}}%
\put(120,#2){\makebox(0,-240)[t]{\small $T$=0}}%
\put(620,#2){\makebox(0,-240)[t]{\small $T$=1}}%
\put(1120,#2){\makebox(0,-240)[t]{\small $T$=2}}%
\put(1620,#2){\makebox(0,-240)[t]{\small $T$=0}}%
\put(2120,#2){\makebox(0,-240)[t]{\small $T$=1}}%
\put(2620,#2){\makebox(0,-240)[t]{\small $T$=2}}%
}
\def\ThreYa#1#2{%
\put(-200,#1){\path(-150,0)(0,0)(-50,20)(-50,-20)(0,0)}%
\put(-450,#1){\rotatebox{90}{\sf \!{#2}}}%
}

\begin{document}

\title{Pentaquark  with diquark correlations\\  in a quark model}
\author{Sachiko Takeuchi%
}\address{%
Japan College of Social Work, Kiyose, Tokyo 204-8555, Japan\\
E-mail: s.takeuchi@jcsw.ac.jp}
\author{Kiyotaka Shimizu}\address{%
Department of Physics, Sophia University, Chiyoda-ku, Tokyo 102-8554, Japan
}

%\date{\today}
\maketitle
\abstracts{
We have investigated uudd$\sbar$ pentaquarks by employing quark models 
with the meson exchange and the effective gluon exchange as qq and q$\qbar$ interactions.
The system for five quarks is dynamically solved;
 two quarks are allowed to have
a diquark-like qq correlation.
It is found that the lowest mass of the pentaquark is about 
1947 -- 2144
MeV. There are parameter sets where
the mass of the lowest positive-parity state become lower than that of the negative-parity states.
Which parity corresponds to the observed peak is still an open question.
Relative distance of two quarks with the attractive interaction is found to be by about 1.2 -- 1.3 times closer than that of the repulsive  one.
The diquark-like quark correlation seems to play an important role 
in the pentaquark systems.
}

\section{Introduction}
Since the experimental discovery of 
the baryon resonance with strangeness
+1, $\Theta$(1540)$^+$, \cite{nakano}
many attempts have been performed to describe the peak theoretically.\cite{oka}
To describe this resonance by using a quark model, one needs at least five quarks, uudd$\sbar$, 
which is called a pentaquark.
A quark model, however, seems to have difficulties
to explain some of the features of this peak.
Namely, 
(1) the  observed mass is rather low, 
(2) the observed width is very narrow, and
(3) there is only one peak is found, especially no $T$=1 peak  nearby.
To reproduce the observed mass, 100 MeV above the KN threshold,
it is preferable to take lowest-mass configuration, $TJ^P$=0\ohalfm.
It is reported, however, that this state will have a very large width,
which contradicts to the observed narrow width.\cite{hosaka}
Other candidate, 0\ohalfp, may have a rather narrow width,
but it might not become as low as the observed one.

The fact that the peak is buried in the NK continuum makes the problem more difficult. As was reported
in this workshop,\cite{latt,qcdsr} the QCD lattice calculation as well as the QCD sum rule approach
found this continuum problem rather serious.
The quark model can deal with this problem by 
introducing the scattering states using the resonating group method.
Also, it is reported\cite{hiyama} that a `bound state'
calculated without such scattering states
becomes  a resonance at almost the same energy when the scattering states are introduced though the levels which couple strongly to 
the nucleon-kaon systems disappear.
In this work, we investigate the pentaquark systems without taking its breaking effect 
as a first step.
Our main aim here is to investigate 
the effects of the qq correlation on the pentaquarks,
which have been mainly treated only by a simple wave function.

We employ several parameter sets for the hamiltonian:
those with the one-boson exchange (OBE), those with 
the one-gluon exchange (OGE), and those with the semirelativistic
 or the nonrelativistic kinetic energy terms.
By employing these various parameter sets, we try to estimate 
the mass of pentaquarks of various quantum numbers with a 
controlled ambiguity.\vspace*{-5pt}

\section{Model}
%\subsection{Hamiltonian}

 The hamiltonian for quarks and anti-quarks is taken as:
 \begin{eqnarray}
H_q &=& \sum_iK_i  +v_0 %\nonumber \\
%&+& 
+\sum_{i<j} \left(\Voge_{ij} + \Vobe_{ij} + V_{\sigma ij} + \Vconf_{ij}\right)
\end{eqnarray}
with
%\\[-14pt]
\begin{eqnarray}
K_i&=& 
\sqrt{m_i^2+p_i^2} 
{\text{\rm ~~(semi-rela)}}~~~~\text{or}~~~~
m_i+{\displaystyle {p_i^2/ 2 m_i}}{\text{\rm ~~(non-rela)}}~.
\end{eqnarray}
The two-body potential term consists of the one-gluon-exchange potential, $\Voge$, the one-PS-meson-exchange
potential, $\Vobe$, the one-$\sigma$-meson-exchange
potential, $V_\sigma$, and the confinement potential, $\Vconf$:
\begin{eqnarray}
\Voge_{ij} &=&\lamilamj{\alpha_s\over 4}\left\{
\left({1\over r_{ij}}\!-\!{e^{-\Lambda_g r_{ij}}\over r_{ij}}\right)
\right.\nonumber\\
&&~~~~~~~~\left.\!-\! \left({\pi\over 2 m_i^2}\!+\!{\pi\over 2 m_j^2}\!+\!{2\pi\over 3 m_im_j}\sigisigj \!\right)
{\Lambda_g^2\over 4\pi}{e^{-\Lambda_g r_{ij}}\over r_{ij}}
\right\}
\\ 
\Vobe_{ij} &=& \frac{1}{3} \frac{g^2}{4 \pi} \frac{m_{\rm m}^2}
{4m_im_j} 
\fifj \sigisigj
%\nonumber \\ && \times 
\left \{ \frac{e^{-m_{\rm m}r_{ij}}}{r_{ij}}
\!-\!\left( \frac{\Lambda_{\rm m}}{m_{\rm m}} \right)^{\!\!2}
\frac{e^{-\Lambda_{\rm m} r_{ij}}}{r_{ij}} \right \}
\\
V_{\sigma~ij} &=& -\frac{g_8^2}{4 \pi} 
\left \{ \frac{e^{-m_{\rm m}r_{ij}}}{r_{ij}}
\!-\!\frac{e^{-\Lambda_{\rm m} r_{ij}}}{r_{ij}} \right \}
\\
{\Vconflam}_{ij} &=& 
-\lamilamj \; \aconf \;r_{ij}~.
\label{eq:conflam}
\end{eqnarray}
In $\Voge$, $\alpha_s$ is the OGE strength, and $\Lambda_g$ is the gluon
form factor.
In $\Vobe$, $g$ is the quark-meson coupling constant:
 $g=g_8$ for  $\pi$, $K$,   and $\eta$,
and $g=g_0$ for $\eta'$ meson. 
The value of $g_8$ can be obtained from the observed 
 nucleon-pion coupling constant, $g_{\pi {\rm NN}}$.\cite{glozm1,FS02}
The term proportional to $\Lambda_{\rm m}^2$
is originally the $\delta$-function term;
the form factor for the meson exchange, $\Lambda_{\rm m}$, is assumed to 
depend on the meson mass $m_{\rm m}$ as
$\Lambda_{\rm m}=\Lambda_0+ \kappa \;m_{\rm m}$.\cite{glozm1,FS02,FST03}

As for the confinement potential for the
five quark system, we replace the factor $\lamilamj$ in eq.\ (\ref{eq:conflam}) 
by its average value as
\begin{eqnarray}
{\Vconfone}_{ij} = {4\over 3} \; \aconf \;r_{ij}~.
\label{eq:conf1}
\end{eqnarray}
This modified confinement potential 
gives the same value for the orbital $(0s)^5$ state as that of the original one.
This replacement enables us to remove all the scattering states
and to investigate only tightly bound states, which will appear
as narrow peaks.
After the coupling to the scattering states with an original confinement 
potential is introduced, some of the states we find will melt away into the continuum.
The situation can be clarified by
evaluating the width, which we will investigate elsewhere.

%
% Table 1
%
\begin{table}[tbp]
\tbl{Five parameter sets.
Each parameter set is denoted by R$\pi$, or Rg$\pi$, etc., 
according to the kinetic energy term and the qq interaction.
Each meson mass is taken to be the observed one, and $m_\sigma$ = 675 MeV.}
{%
\def\SPA{\phantom{0}}%
\def\SPB{\phantom{$-$0}}%
\def\SPC{\phantom{.0}}%
\tabcolsep=0.3mm%
%\footnotesize
\begin{tabular}{l@{~}ccccccccccccccc}
\toprule
Model& qq int.& $m_{\rm u}$ &$m_{\rm s}$ 
& $\alpha_s$ & $\Lambda_g$ & $\FRAC{$g^2_8$}{$4\pi$}$ 
&$(\!\FRAC{$g_0$}{$g_8$})^2$& $\Lambda_0$ &$\kappa$ &$\aconf$ & $V_0$
\\
ID&&\MeV&\MeV&&\footnotesize [fm\!\!$^{-1}$]&&&\footnotesize [fm\!\!$^{-1}$]&&\footnotesize [MeV\!/fm] &\MeV \\ \midrule
R$\pi^\dag$  & $\pi$ $\sigma$ $\eta$& 313 & 530 & 0 & - & 0.69\SPA &0& 1.81 &0.92 & 170\SPC &$-$378.3 \\
Rg$\pi$  & \OGE\ $\pi$ $\sigma$ $\eta$ $\eta'$& 340 & 560 & 0.35 & 3 & 0.69\SPA & 1 & 1.81 & 0.92  & 172.4 & $-$381.7 & \\
%N$\pi$  & $\pi$ $\eta$ $\eta'$ &330 & 404 &  0 & - & 0.85\SPA &2.3& 8.23 & 1.3\SPA &-& \SPA{}70\SPC & $-$173.7 & \cite{FST03}\\
Ng$\pi^\ddag$  & \OGE\ $\pi$ $\sigma$ $\eta$ &313 & 550 &  0.35 & 5 & 0.592 &0& 2.87 & 0.81 & 172.4 & $-$453\SPC\\
Ng  & \OGE &313 & 680 &1.72& 3 & 0 & - & - & -  & 172.4 & $-$345.5 &\\ %\midrule
Graz$^\S$ & $\pi$ $\eta$ $\eta'$&340 & 500 &0 & - & 0.67\SPA &1.34& 2.87 &0.81 &172.4 & $-$416\SPC &\\ \bottomrule
\multicolumn{5}{l}{\rule{0mm}{3mm}$^\dag$Ref.~\refcite{FST03},  $^\ddag$Ref.~\refcite{FS02}, and $^\S$Ref.~\refcite{glozm1}.}
\end{tabular}\label{tbl:QMparam}}
\vspace*{-13pt}
\end{table}
%
% Table 2
%
\begin{table}[tb]
\tbl{$S$-wave and $P$-wave quark pairs}
%\begin{center}
%\renewcommand{\arraystretch}{1.05}
{
\begin{tabular}{ccccc}
\toprule
& $T$ & $S$ & $C$ & $(-)^\ell$ \\ \midrule
$\alpha$   & 0 & 0 & $\threebar$ &+\\ 
$\beta$    & 0 & 1 & ${\bold 6}$ &+\\ 
$\gamma$ & 1& 0 & ${\bold 6}$ &+\\ 
$\delta$  &1 &1 & $\threebar$ &+\\ 
%\midrule
\bottomrule
\end{tabular}
~~
\begin{tabular}{ccccc}
\toprule
& $T$ & $S$ & $C$ & $(-)^\ell$ \\ \midrule
$\smash{\talpha}$   & 0 & 0 & ${\bold 6}$ &$-$\\ 
$\smash{\tbeta}$    & 0 & 1 & $\threebar$ &$-$\\ 
$\smash{\tgamma}$ & 1& 0 & $\threebar$ &$-$\\ 
$\smash{\tdelta}$  &1 &1 & ${\bold 6}$&$-$\\  \bottomrule
\end{tabular}
\label{tbl:diquarks}
}
%\end{center}
\end{table}

We have employed four kinds of parameter sets: R$\pi$, Rg$\pi$,
Ng$\pi$ and Ng (Table \ref{tbl:QMparam}).
R stands for the parameter sets with the semirelativistic kinetic energy term, while N stands for those with the nonrelativistic
one.
The parameter sets with OBE [OGE] are denoted by the name with $\pi$ [g].
% For reference, w
We also perform the calculation with the parameter set 
%given by Ref.~\refcite{glozm1}.
given by Graz group.\cite{glozm1}
\bigskip

The wave function we employ  is written as:
\begin{eqnarray}
\lefteqn{\psi_{TSL}(\vecxi_A,\vecxi_B,\veceta,\vecR)=\sum_{i,j,n,m,\omega,\omega',\lambda} 
 c_{ijnm}^{\omega \omega' \lambda}\;
{\mathcal{A}}_{q^4}}&&\nonumber \\%[-1.5ex]
&&~~\times
\phi_{\rm q^2}(\omega,\vecxi_A;u_i)\;\phi_{\rm q^2}(\omega',\vecxi_B;u_j)\;
\psi(\lambda,\veceta;v_n)\Big|_{L}\;\chi_{\sbar}(\vecR;w_m)~,
\end{eqnarray}
where ${\mathcal{A}}_{q^4}$
is the antisymmetrization operator over the four ud-quarks, and
$\vecxi_A$, $\vecxi_B$, $\veceta$ and $\vecR$
are the internal coordinates defined as:
\begin{eqnarray}
\vecxi_A &=& \vecr_1-\vecr_2 ~~{\rm and}~~
\vecxi_B = \vecr_3-\vecr_4~, \\
\veceta &=& (\vecr_1+\vecr_2 -\vecr_3-\vecr_4 )/2~,\\
\vecR &=& (\vecr_1+\vecr_2 +\vecr_3+\vecr_4)/4-\vecr_{\sbar}~.
\end{eqnarray}
$\phi_{\rm q^2}(\omega,\vecxi;u)$ is the wave function for two quarks with quantum number $\omega$,
which is one of $\alpha,\beta,\cdots,\tdelta$ listed in 
Table \ref{tbl:diquarks}, 
with size parameter $u$:
\begin{eqnarray}
\phi_{\rm q^2}(\omega,\vecxi;u)&=& 
\left\{ \begin{array}{c} \varphi([11]_{STC}) \\ \varphi([2]_{STC}) \;\vecxi  \\ \end{array} \right\}
\exp\left[-\dfrac{\xi^2}{4 u^2}\right] 
~~~\left\{ \begin{array}{c} (\ell=0)\\ (\ell=1)\\ \end{array} \right\}~.
\end{eqnarray}
The relative wave function between two qq pairs, $\psi(\lambda,\veceta;v)$ and 
the wave function between four quarks and $\sbar$ quark, $\chi_{\sbar}(\vecR;w)$, are
taken as:
\begin{eqnarray}
\psi(\lambda,\veceta;v)&=&
\left\{ \begin{array}{c} 1\\ \veceta\\ \end{array} \right\}
\exp\left[-\dfrac{\eta^2}{2 v^2}\right] 
~~~\left\{ \begin{array}{c} (\lambda=0)\\ (\lambda=1)\\ \end{array} \right\}
\\[1.2ex]
\chi_{\sbar}(\vecR;w)
&=&\exp\left[-\dfrac{2R^2}{5 w^2}\right].
\end{eqnarray}
%The four-quark  wave function is taken as follows.
For the negative-parity pentaquarks ($L$=0), we use all possible $\ell$=$\ell'$=$\lambda$=0
states.
For the positive-parity pentaquarks ($L$=1),
we use  the states where one of  $\ell$ $\ell'$ and $\lambda$ is equal to 1.
The gaussian expansions are taken as geometrical series:
$u_{i+1}/u_i$ = $v_{n+1}/v_n$ = $w_{m+1}/w_m$ = 2.
We take 6 points for $u$ (0.035--1.12 or 0.04--1.28fm),
4 points for $v$ (0.1--0.8fm), and 3 points for $w$ (0.2--0.8fm).
Since we use a variational method, the obtained masses are the upper-limit.
They, however, converge rapidly; the mass may reduce more, but probably only by several MeV.\vspace*{-10pt}

%\newpage
\section{Results and discussions}

The mass of the q$\qbar$, q$^3$,  and q$^4\qbar$ systems are shown in Table \ref{tbl:mbmass}.
Parts of these baryon masses were given in refs.\ \refcite{FS02} and \refcite{FST03}.

It is very difficult for a constituent quark model to describe the Goldstone bosons.
Also, it is hard to justify the models with the kaon-exchange interaction between
quarks to describe a kaon.
We use the K$^*$ mass as a reference 
for the pentaquarks.

Contrary to the q$\qbar$ systems, 
we have more satisfactory results for the q$^3$ baryons. The mass spectrum of the $S$-wave ground states is well reproduced.
Moreover, since the chiral quark models have a mechanism to lower the mass of the Roper resonance than that of the 
negative-parity excited nucleons, the excited baryon mass spectrum can also be reproduced.
On the other hand, in the OGE quark model picture, the Roper mass is considered to reduce by introducing the pion-cloud effect, which should be 
taken into account separately.
Though it is interesting to see whether the Roper resonance has a pentaquark component,\cite{JW,KL} that is out of scope of our present work.
Parameter sets Rg$\pi$ and Ng underestimate the negative-parity excited nucleon mass 
by about 70 MeV and 90 MeV, respectively.
When we discuss the positive-parity pentaquarks
by these models, we will have to take this underestimate into account.

\begin{table}[btp]
\tbl{Masses of meson, baryon, and pentaquark, P($TJ^P$), in MeV.}
{\tabcolsep=1mm\label{tbl:mbmass}
\def\SPC{\phantom{.0}}
\begin{tabular}{l@{~~}cccc@{~~~~}cc@{~~~}cc@{~~~}cc@{~~~}cccc}
\toprule
         & N &$\Delta$&K$^*$&NK$^*$& 
\multicolumn{2}{c}{P(0\ohalfm)} & \multicolumn{2}{c}{P(0\thalfm)} 
& \multicolumn{2}{c}{P(1\ohalfm)} & \multicolumn{2}{c}{P(0\ohalfp)$^\ddag$}\\[0.2ex]
&&&&                               & $m_P$&$m_P'$& $m_P$&$m_P'$& $m_P$&$m_P'$& $m_P$&$m_P'$ \\ \midrule
R$\pi$   & 941 & 1261 & 979 & 1921 & 2109 & 2054 & 2141 & 2083 & 2143 & 2083 & 2165 & 2045 \\
Rg$\pi$  & 938 & 1232 & 931 & 1869 & 1985 & 1947 & 2064 & 2018 & 2078 & 2021 & 2129 & 2006 \\
Ng$\pi$  & 936 & 1232 & 814 & 1846 & 2029 & 1996 & 2106 & 2066 & 2106 & 2061 & 2321 & 2144 \\
Ng       & 938 & 1232 & 814 & 1846 & 1966 & 1971 & 2153 & 2144 & 2170 & 2145 & 2345 & 2197 \\
Graz     & 937 & 1239 & 927 & 1864 & 2231 & 2160 & 2240 & 2168 & 2251 & 2173 & 2248 & 2120 \\ \midrule
Exp.$^\dag$& 939 & 1232 & 892 & 1831 & 
\multicolumn{8}{c}{1540}\\ \bottomrule
\multicolumn{13}{l}{\rule{0cm}{3mm}$^\dag$Ref.\ \refcite{pdg}. $^\ddag$Since  the interaction we use is central,
0\ohalfp\ and 0\smash{${3\over 2}^+$} are degenerated.}
\\
\end{tabular}\par
}
\vspace*{-8pt}
\end{table}

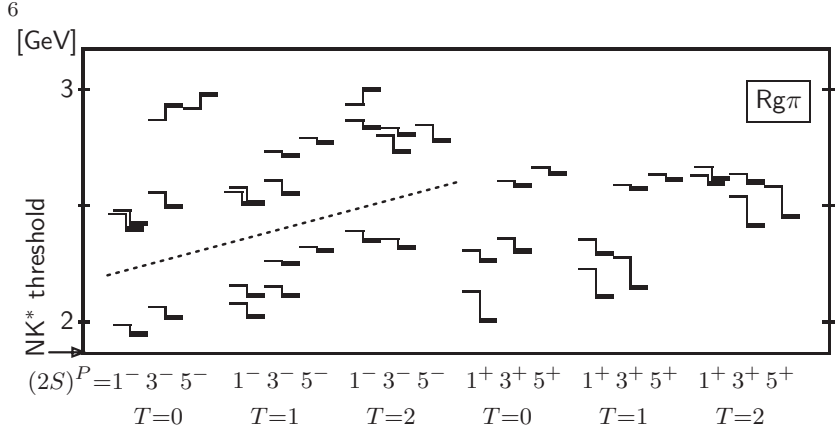
\begin{figure}[bt]
%\epsfxsize=10cm   %width of figure - will enlarge/reduce the figures
%\epsfbox{fig3.eps}
%\figurebox{2cm}{3cm}{} %to have a box alone 
%\includegraphics*[scale=0.5]{s_takeuchi_fig.eps}
%\centerline{\epsfxsize=4.1in\epsfbox{procs-fig1.eps}}   
{%
\unitlength=0.088pt%
\thicklines%
\begin{center}
\rotatebox{0}{%
\def\thre{1869}%
%\def\threD{2067}
%\begin{picture}(3200,2000)(0,2000)
\begin{picture}(2600,1500)(0,1600)
\Waku{{Rg$\pi$}}{\thre}{1300}
\ThreYa{\thre}{NK$^*$ threshold}
%\ThreYa{\threD}{K$\Delta$ threshold}
\put(-225,3200){\makebox(0,0)[r]{\sf [GeV]}}
%\put(-200,2000){\framebox(3200,1200)}
\put(-225,2000){\line(1,0){50}}
\put(-300,1970){\sf 2}
\put(-225,3000){\line(1,0){50}}
\put(-300,2970){\sf 3}
\put(-225,2500){\line(1,0){50}}
\put(2975,2000){\line(1,0){50}}
\put(2975,2500){\line(1,0){50}}
\put(2975,3000){\line(1,0){50}}
%
% 0 and 2 hbar omega
%\put(-100,2200){\line(4,1){1500}}
\dottedline{40}(-100,2200)(1400,2600)
%\put(-100,2200){\line(4,1){1500}}
%R-all (weaker OGE)
\Msx{0}{0}{1}{2463}{2399}{-1}{64}
\Ms{0}{1}{1}{1985}{1947}{-1}{38}
\Ms{0}{1}{3}{2064}{2018}{-1}{46}
\Ms{0}{2}{3}{2868}{2931}{1}{63}
\Ms{0}{2}{5}{2917}{2980}{1}{63}
\Ms{1}{0}{1}{2078}{2021}{-1}{57}
\Ms{1}{1}{1}{2157}{2112}{-1}{45}
\Ms{1}{1}{3}{2153}{2112}{-1}{41}
\Ms{1}{2}{3}{2260}{2252}{-1}{8}
\Ms{1}{2}{5}{2321}{2308}{-1}{13}
\Ms{2}{0}{1}{2934}{2998}{1}{64}
\Ms{2}{1}{1}{2391}{2348}{-1}{43}
\Ms{2}{1}{3}{2357}{2321}{-1}{36}
\Msx{2}{2}{3}{2801}{2735}{-1}{66}
\Ms{2}{2}{5}{2847}{2778}{-1}{69}
\Mp{0}{0}{1}{2129}{2006}{-1}{123}
\Mp{0}{1}{1}{2307}{2263}{-1}{44}
\Mp{0}{1}{3}{2359}{2305}{-1}{54}
\Mp{0}{2}{3}{2605}{2585}{-1}{20}
\Mp{0}{2}{5}{2663}{2639}{-1}{25}
\Mp{1}{0}{1}{2354}{2295}{-1}{59}
\Mp{1}{1}{1}{2226}{2107}{-1}{119}
\Mp{1}{1}{3}{2276}{2148}{-1}{128}
\Mp{1}{2}{3}{2588}{2573}{-1}{15}
\Mp{1}{2}{5}{2635}{2614}{-1}{21}
\Mp{2}{0}{1}{2666}{2618}{-1}{48}
\Mpx{2}{1}{1}{2630}{2595}{-1}{35}
\Mp{2}{1}{3}{2636}{2601}{-1}{35}
\Mp{2}{2}{3}{2538}{2415}{-1}{123}
\Mp{2}{2}{5}{2581}{2454}{-1}{127}
% next
%\Ms{0}{1}{1}{1985}{1947}{-1}{38}
\Ms{0}{1}{1}{2478}{2424}{-1}{54}
%
%
%\Ms{0}{1}{3}{2064}{2018}{-1}{46}
\Ms{0}{1}{3}{2556}{2497}{-1}{60}
%
%
%\Ms{1}{0}{1}{2078}{2022}{-1}{57}
\Ms{1}{0}{1}{2577}{2512}{-1}{65}
%
%
%\Ms{1}{1}{1}{2157}{2112}{-1}{45}
\Msx{1}{1}{1}{2558}{2509}{-1}{49}
%
%
%\Ms{1}{1}{3}{2153}{2112}{-1}{41}
\Ms{1}{1}{3}{2608}{2551}{-1}{57}
%
%
%\Ms{1}{2}{3}{2260}{2252}{-1}{8}
\Ms{1}{2}{3}{2732}{2717}{-1}{15}
%
%
%\Ms{1}{2}{5}{2321}{2307}{-1}{13}
\Ms{1}{2}{5}{2790}{2773}{-1}{18}
%
%
%\Ms{2}{1}{1}{2391}{2348}{-1}{43}
\Ms{2}{1}{1}{2865}{2837}{-1}{28}
%
%
%\Ms{2}{1}{3}{2357}{2320}{-1}{36}
\Ms{2}{1}{3}{2833}{2807}{-1}{26}
\end{picture}
}%
\end{center}
}%
\caption{Pentaquark mass spectrum obtained by the Rg$\pi$ parameter set. \label{fig:pentamass}}
\vspace*{-13pt}
\end{figure}

Now we discuss the system of the pentaquarks.
There is no bound state when we use the original confinement, eq.\ (\ref{eq:conflam}).
In Table \ref{tbl:mbmass}, we show
the masses of the pentaquarks with $\Vconfone$, $m_P$, 
and the mass with the correction from the confinement potential
evaluated by the wave function corresponding to $m_P$:\vspace*{-5pt}
\begin{equation}
m_P' = m_P+ \bra {\Vconflam}_{ij}  \ket - \bra {\Vconfone}_{ij} \ket~.
\label{eq:confdiff}
\end{equation}
In Figure \ref{fig:pentamass}, we plot these $m_P$ (thin bars)
and $m_P'$ (thick bars) for Rg$\pi$.
Among the q$^4$ $S$-wave systems, five spin-isospin states can couple to the orbital $(0s)^4$ configuration:
 ($TS$)=(01), (10), (11), (12), and (21).\cite{ST04}
%The negative-parity pentaquarks under the dotted line in Figure \ref
%{fig:pentamass} corresponds roughly to those which include the above states.
The negative-parity pentaquarks which have a large component of these q$^4$ states 
 correspond to the levels under the dotted line in Figure \ref{fig:pentamass}.
The mass difference between them and other states is about 
  several hundred MeV.
Among them, the
 ($TS$) = (01) and (10) states are the lowest two states, which are essentially degenerated: {\it i.e.}, 
($TS$)$J^P$=(01)\ohalfm, (01)\thalfm, and (10)\ohalfm.
As OGE becomes stronger, (01)\ohalfm\ goes down. For example, the splitting between (01)\ohalfm\ and  \thalfm\
is 71 MeV in Rg$\pi$ whereas it is 32 MeV in $R\pi$, or 9 MeV in Graz.
It becomes 174 MeV in Ng, where all the hyperfine splitting comes from OGE.
The remaining two levels, however, still stay close to each other.

There is not such a large separation in the mass spectrum of the positive parity pentaquarks, reflecting the fact that all of the spin-isospin states can couple to 
 the orbital $(0s)^40p$ state.
The interaction, however, makes one of the states very low: {\it i.e.}, (00)\ohalfp\ (and \smash{${3\over 2}^+$} because the interaction
 is central in the present work).
It can actually be as low as the negative-parity states.

The masses are still much higher than the observed peak, and depend on the parameters.
There is a large ambiguity in the zero-point energy\cite{TS04} as well as in the interaction.
For example, we should include the instanton induced interaction,\cite{shinozaki} which is the source
of the $\eta$-$\eta'$ mass difference.
It seems, however, that
the relative positions of the levels 
do not change much. We argue that one of 
the above mentioned levels is observed as the peak.

Because the $TJ^P$=0\ohalfm\ and 1\ohalfm\ 
states couple to the NK state strongly and 1\thalfm\ couples to
$\Delta$K,
it is unlikely that they appear as narrow peaks.
It seems that 0\thalfm\ is a good candidate for the observed peak.\cite{TS04}
Unfortunately, which of the above 
0\thalfm\ and 0\ohalfp\ 
states is most likely seen is  still an open question.
The 0\thalfm\ is lower than the other in  
Ng$\pi$ and Ng, and also in the Rg$\pi$ parameter set
if its underestimate of the $P$-wave baryon mass is taking into account.
 On the other hand, 
 the 0\ohalfp\ state is lower than the other 
 in the semirelativistic chiral models: R$\pi$ and Graz.
In all the cases, however, the mass difference between these two states
is not large.
%\bigskip
%\subsection{Full calculation}
%\clearpage
%\subsection{Roles of the qq correlation}

\begin{table}[bt]
\tbl{Values of $\OFS$
for each q$^4$ state.  }
{\label{tbl:matele_chiral_dq}
\def\SPA{\phantom{0}}
\def\SPB{\phantom{$-$0}}
\def\SPC{\phantom{.0}}
\begin{tabular}{llr@{~(}r@{~)~~}l@{~~~~~}lll}
\toprule
%&&&&&&&\multicolumn{4}{c}{\# of quark pairs of $(T_2S_2)$}\\
&($T,S$) 
&\multicolumn{2}{c}{$\OFS$} & ~~~~diff & \multicolumn{2}{c}{Full (R$\pi$) [MeV]}
\\ \midrule
Parity $-$ 
&(01),(10) & $-$10  &$-$16 &\raisebox{-.7em}[0pt][0pt]{$\ket$ \SPA{}4 (4)} &2131 &\raisebox{-.7em}[0pt][0pt]{$\ket$ 100}\\ 
&(11) & $-$6  &$-$12 &\raisebox{-.7em}[0pt][0pt]{$\ket$ \SPA{}8  (8)} & 2231&\raisebox{-.7em}[0pt][0pt]{$\ket$ 187}\\ 
&(12),(21) &  2  &$-$4 && 2418 \\ \midrule
Parity $+$ &
(00) & $-$30  &$-$30 &\raisebox{-.7em}[0pt][0pt]{$\ket$ \SPA{}8 (8)}  &2178 &\raisebox{-.7em}[0pt][0pt]{$\ket$ 164}\\ 
&(11) & $-$22  &$-$22 &\raisebox{-.7em}[0pt][0pt]{$\ket$ 12 (6)} &2342 &\raisebox{-.7em}[0pt][0pt]{$\ket$ 123}\\ 
&(01) ,(10) & $-$10  &$-$16 && 2465\\ 
\bottomrule
\end{tabular}
}\vspace*{-5pt}
\end{table}

Except for the confinement force, all the interaction terms
are short-ranged in the quark model.
Thus, when the two-quark correlation is introduced in the model,
quark pairs where the interaction 
is attractive come closer while
those with repulsion tend to stay apart from each other.
Then an attractive pair may behave like a single particle, called a diquark.

In the chiral quark model with a simple gaussian wave function, the matrix element of the spin-isospin operator,
$\OFS$, is proportional to the hyperfine splitting (Table \ref{tbl:matele_chiral_dq}):
\begin{eqnarray}
\OFS &=&-\bra[f]_{TS}|\sum_{i<j}(\tau_i\cdot\tau_j)(\sigma_i\cdot\sigma_j)|[f]_{TS}\ket~. %\\
%\OCS &=&-\bra[f]_{CS}|\sum_{i<j}(\lambda_i\cdot\lambda_j)(\sigma_i\cdot\sigma_j)|[f]_{CS}\ket
\end{eqnarray}

On the other hand, suppose one takes a diquark-model picture, only 
the pairs between which the interaction is attractive have to be considered.
The expectation values of $\OFS$ 
in this picture are listed
 in the parentheses in Table \ref{tbl:matele_chiral_dq}
 alongside of the original matrix elements.
The `mass difference' between these states is also 
shown in the column under `diff'.
The ratio of the mass differences of the positive-parity states is 8/12 in the shell-model picture while it is 8/6 in the diquark-like picture.
The ratio obtained from the averaged masses by our full calculation is found to support  the 
diquark-like picture.
The qq correlation plays an important role in the pentaquarks.
%\bigskip

\begin{table}[bt]
\tbl{Size 
of the quark pairs of each $(T_2S_2)$ spin-isospin state
in the ($TS$)$J^P$=(01)$\smash{{1\over 2}}^-$ and (00)$\smash{{1\over 2}}^+$ pentaquarks as well as  that in the nucleon.}
{%
\label{tbl:rr}%
\begin{tabular}{c@{~~~~~}rrrr@{~~~~~~~}rrrr}
\toprule
&\multicolumn{4}{c}{$(T_2S_2)$ pair (R$\pi$)}
&\multicolumn{4}{c}{$(T_2S_2)$ pair (Rg$\pi$)}\\
($TS$)$J^P$  &(00) & (01) & (10) & (11)  &(00) & (01) & (10) & (11)  
\\ \midrule
 (01)1/2$-$ 
& 0.53 & 0.70 & 0.68 & 0.62 & 0.56 & 0.69 & 0.68 & 0.64
\\
 (00)1/2+
& 0.56 & 0.69 & 0.69 & 0.61 & 0.61 & 0.74 & 0.74 & 0.68
\\
N 
& 0.50 & 0.65 & 0.65 & 0.56 & 0.55 & 0.76 &0.76 & 0.62
\\
\bottomrule
\end{tabular}
}\vspace*{-5pt}
\end{table}

More direct approach to see the importance of the qq correlation is to look into the size of the quark pairs, $\sqrt{\bra r^2 \ket}$.
In Table \ref{tbl:rr} we show the size of quark pair
for each ($T_2S_2$) state in the lowest pentaquarks.
Their size is large when the interaction is repulsive while it becomes small for the attractive
pairs.
The ratio is about 1.2 -- 1.3.
The degree of the qq correlation in the pentaquarks seems  similar to that in the nucleon.\vspace*{-10pt}

\section{Summary}

We have investigated uudd$\sbar$ pentaquarks by employing quark models 
with the meson exchange and the effective gluon exchange as qq and q$\qbar$ interactions.
The system for five quarks is dynamically solved;
two quarks are allowed to have
a diquark-like qq correlation.

The present work indicates that
the \smash{$TJ^P$}=0\thalfm, 0\ohalfp\ pentaquark states
can be almost as low as the 0\ohalfm\ state, which has
been assigned to the observed peak, but expected to have a large width.
Both of the 0\thalfm\ and 0\ohalfp\ states
are considered to have a narrow width.
Which parity should correspond to the observed peak is still an open question.
Relative distance of two quarks with the attractive interaction is found to be by about 1.2 -- 1.3 times closer than that of the repulsive  one.
The diquark-like quark correlation seems to play an important role 
in the pentaquark systems.
The  1\ohalfm\  state is also found to be low.
Like the 0\ohalfm\  state, however, 
it couples to the NK states strongly. 
This may be the reason
why there is no peak in the $T$=1 channel.

As for the absolute mass, our estimate is still more than 400 MeV
higher than the observed one.
We consider the extra attraction may come from other 
qq interactions as well as from the 
ambiguous zero-point energy.
The width and the resonant energy should
be investigated by including the coupling to the baryon meson asymptotic states, which is underway.

%\section*{acknowledgements}
\vfill

This work is supported in part by a Grant-in-Aid for Scientific Research
from JSPS (No.\ 15540289).%\vspace*{-10pt}

\end{document}